\documentclass[11pt]{article}
\usepackage{graphicx,latexsym,amsfonts,graphicx}
\usepackage{epstopdf}
\usepackage[usenames]{color}
\usepackage{calc}
\usepackage{verbatim}
\usepackage[mathcal]{euscript} 
\usepackage{subfig}
\usepackage{multicol}
\usepackage{geometry}
\usepackage{amssymb}
\usepackage{mathtools} %
\usepackage{url}

\newcommand{\R}{\ensuremath{\mathbb{R}}}

\newcommand{\be}{\begin{eqnarray}}
\newcommand{\ee}{\end{eqnarray}}

\newcommand{\EE}{\mathbb{E}}

\newcommand{\indep}{\perp \!\!\! \perp}
\newcommand{\rhot}{\tilde{\rho}}

\title{The  hierarchical barycenter: conditional probability simulation with structured and unobserved covariates}

\author{Esteban G. Tabak \and Giulio Trigila \and Wenjun Zhao}
\providecommand{\keywords}[1]{\textbf{\textit{Index terms---}} #1}

\date{Sept 29, 2024}

\begin{document}

\maketitle

\begin{abstract}
This paper presents a new method for conditional probability density simulation.The method is design to work with unstructured data set when data are not characterized by the same covariates yet share common information. Specific examples considered in the text are relative to two main classes: homogeneous data characterized by samples with missing value for the covariates and data set divided in two or more groups characterized by covariates that are only partially overlapping. The methodology is based on the mathematical theory of optimal transport extending the barycenter problem to the newly defined hierarchical barycenter problem. A newly, data driven, numerical procedure for the solution of the hierarchical barycenter problem is proposed and its advantages, over the use of classical barycenter, are illustrated on synthetic and real world data sets. 
   
\end{abstract}

\section{Introduction}

This article develops a data-based methodology for simulating conditional distributions, when some of the covariates are only observed or even defined  in overlapping subsets of the data. Consider as an example a recommendation system where users assign a rate $x$ to objects of different classes, such as movies, books and bicycles. How could one estimate, based on data, the rate that a particular user will assign to a particular object?

Every individual rate $x_i$ has a few qualifiers attached. Some may relate to the user (e.g. age, national origin), some to the object (a book's number of pages, a bicycle's color), some to environmental factors (day of the week, weather, location). We will group these qualifiers into a factor $z_i$ with entries $\{z_i^l\}$. Our goal is, given new factor values $z_*$, estimate  the corresponding conditional density $\rho(x|z_*)$ or simulate it by drawing samples from it.

If the set of factors $\{z^l\}$ were common to all observations $x_i$, we could use existing tools, such as kernel conditional density estimation \cite{fan1996estimation,de2003conditional}, or conditional density simulation through the distributional barycenter problem, a methodology  developed by the authors in prior work \cite{tabak2020conditional,tabak2022distributional}. 
The $z$-variable in our problem, however, is not a regular vector, as it is not globally defined in a uniform way. The $z^l$ corresponding to number of pages, for instance, is defined for books but not for bicycles.

We could reduce this problem to the regular one by dividing it into sub-problems: one for movies, one for toys, and so on. Yet each of these sub-problems has a much smaller number of observations available than their aggregation, yielding less samples to base our estimation on. This sample-size problem is exponentiated by the fact that  the tree of factors $z$ may branch further, with only science books, for instance, classifiable into physics, zoology and so on. Yet each sub-problem could inform the others: the user's age, for instance, may have a consistent effect across the rates, and choices made on objects of one type could well inform those to be made among objects of other types. 

Even though we have used a recommendation system for illustration, the setting described applies much more broadly. The variables $x$ can quantify, for instance, the results of a medical treatment, where the condition being treated brings in its own qualifiers, such as body temperature for a cold or blood glucose content for diabetes, and each drug administered may be further qualified by the corresponding specification and dosage. A similar problem appears whenever one would like to bring together different data bases: the positive effect of their aggregation on the sample size comes at the price of having different sets of covariates available for each. Prediction with missing data, where the unavailability of different components of the vector $z$ of predictors for each observation is due to lack of knowledge, not of definition, also fits in a similar framework.

This article proposes a methodology, the hierarchical barycenter, to address this broad category of data problems, extending the optimal transport-based methodology in \cite{tabak2020conditional} to covariates with a much more general structure.

For the sake of clarity, we will focus on two main categories of problems for which the hierarchical barycenter is particularly well suited:
\begin{enumerate}
\item Structured cofactors: the data in the training set have no missing values, they are however divided into groups characterized by different cofactors (e.g. data relative to books and to movies). 

\item Missing data: the training set shares a common set of cofactors, yet only some of these are available for each data point. 
\end{enumerate}

The paper is structured as follows. Section \ref{sec:Formulation} describes the hierarchical barycenter problem within the frame of the theory of optimal transport \cite{santambrogio2015optimal}. Section \ref{sec:Alg} describes the details of the implementation of the hierarchical barycenter that we then use  in Section \ref{sec:Num} to  analyze synthetic and real world data relative to mineral bone density \cite{hastie2009elements,sugiyama2010conditional}.

\section{Formulation}\label{sec:Formulation}

\subsection{The distributional barycenter problem}\label{sec:DB}
Before addressing the structure of the factors $z$, this section   summarizes the distributional barycenter problem \cite{tabak2022distributional}, an extension of the Wasserstein barycenter described in \cite{AguehBarycenters}, and its use to perform conditional density simulation. This problem can be posed as follows: given a set of $N$ sample pairs $\{x_i, z_i\}$ --the observations-- drawn from an unknown joint distribution
$$ \pi(x, z) = \gamma(z) \rho(x|z), $$
simulate
$$ \rho\left(x | z_*\right) $$
for any target value $z_*$ of $z$, i.e. extract samples $\{x^*_j\}$ thereof. 

The distributional barycenter removes from the $\{x_i\}$ all the variability that the $\{z_i\}$ can explain --and only that variability--  transforming them into new variables $\{y_i\}$ that are independent of $z$. If one then brings back to the $\{y_i\}$ the variability that $z = z_*$ would entail, the resulting $\{x^*_i\}$ are the samples of $\rho\left(x | z_*\right)$ sought.

Describing this procedure in more detail, one first seeks a map $y = T(x, z)$ with $x, y\in \R^{d_{x}}$ and the factors $z \in \R^{d_{z}}$ acting as parameters. Removing from $x$ all variability that $z$ can explain translates into the condition that the resulting $y$ must be independent of $z$. Not removing any additional variability from $x$ can be implemented through the requirement that the map $T$ deform $x$ as little as possible. Then the distributional barycenter problem reads
\begin{equation}
 \min_{y = T(x, z)} C(x, y) \quad \hbox{s.t.} \quad y \indep z, 
 \label{Barycenter}
\end{equation}
where the symbol ``$\indep$'' stands for independence, and $C(x, y)$ is a measure of the deformation incurred by transforming the $x$ into $y$'s. In an optimal transport framework, $C$ adopts the form of the expected value of a pairwise deformation cost $c$,
\begin{equation}
 C(x, y) = \EE[c(x, y)],
 \label{cost}
\end{equation}
where a typical choice for $c$ in normed spaces is the squared norm
$$ c(x, y) = \frac{1}{2} \|y - x\|^2. $$
Once $T$ has been found, we draw samples $\{x^*_i\} \sim \rho\left(x | z_*\right)$ for any target $z_*$ through
\begin{equation}\label{eq:prediction}
x^*_i = T^{-1}\left(y_i, z_*\right), \quad \hbox{where} \quad  y_i = T\left(x_i, z_i\right). 
\end{equation}
The formulas in (\ref{eq:prediction}) carry out the program described above, removing first from the $\{x_i\}$ the variability attributable to the corresponding $\{z_i\}$, and then replacing it by the variability entailed by the target $z_*$.

\subsection{Extension to hierarchical covariates}\label{sec:extension}

Applying the procedure just described to our problem requires understanding what it means for $y$ to be independent of $z$ when the latter has components that are defined only for a subset of the data. The simplest setting has a single variable $z^1$ that is known for some of the $\{x_i\}$ and not for the others. Let $I_k$ be the set of observations $\{i\}$ where $z^1$ is known and $I_u$ its complement. If one simply removes the variability in $x$ attributable to $z^1$ from the $\{x_i, \ i \in I_k\}$ while leaving the $\{x_i, \ i \in I_u\}$ untouched, the resulting $\{y_i\}$ will be divided into two groups, one with reduced variability and the other not. This spurious source of variability in $y$ can be explained away by adding a new, binary factor $z^0 \in \{k, u\}$, discriminating observations in $I_k$ from those in $I_u$.  

It follows that we should use in this case a factor $z$ structured as a tree, with $z^0$ partitioning the root into two branches, one with and one without the covariate $z^1$. Then the independence between $y$ and $z$ acquires a clear meaning: $y$ must be independent of $z^0$,  and those $y$ with $z^0 = k$ must be independent of $z^1$. Notice that, if the availability of $z^1$ was caused by a hidden binary confounder that could also affect the distribution of the $x$, this additional source of variability is also taken into consideration by the inclusion of $z^0$ as a factor. On the other hand, if this confounder was not hidden but interpretable --we can count the number of pages $z^1$ of a book but not those of a bicycle-- then $z^0$ would have been already included as a factor among the $z^l$ --in this example, through the type of object being rated.

In the general case, the structure of $z$ may be not that of a tree but of a more general graph, for instance when different subgroups of objects have partially overlapping factors, such as color, size and age. The general rule is that, for any observation $x_i$, the subset of the factors $\{z_i^l\}$ available should be fully determined by some of those factors themselves, either in an interpretable fashion --as with weight being an available factor for luggage-- or through a factor added explicitly to account for possibly missing observations. With such \emph{covariate extension}, the problem formulation reduces to a regular barycenter problem. 

The factors $z_i^{l}$ associated to different data sets need not be of the same type or dimensionality, as for instance, books and movies may have a different number of covariates. To keep the notation simple, we will omit to specify the type and dimensionality of each factor $z_i^{l}$ . 

\subsection{Enforcing independence between $y$ and $z$}

The objective function of the barycenter problem (\ref{Barycenter}) has two components: the cost $C(x, y)$ to minimize and the constraint that $y$ and $z$ should be independent. Translating the cost $C$ to a sample-based formulation is straightforward, particularly in an optimal transport setting, where when the joint distribution $\pi(x, z)$ is only known through $N$ sample pairs $\left\{x_i, z_i\right\}$, we can replace expectation with empirical mean:
\begin{equation}
 C(x, y) = \EE[c(x, y)] \rightarrow\ \frac{1}{N} \sum_i c\left(x_i, y_i\right), \quad y_i = T\left(x_i, z_i\right).
\end{equation}
In order to complete the problem's formulation, we also need to implement the independence condition in a sample-friendly way.

There are a number of ways to write down the condition that two variables $y$ and $z$ be independent, i.e. that their joint distribution factorizes:
\begin{equation}
\pi(y, z) = \rho(y) \gamma(z). 
\label{factorization}
\end{equation}
Some choices are:
\begin{enumerate}

\item A weak formulation of (\ref{factorization}) in terms of measurable test functions:
$$ \forall F(y, z), \ \int \left[F(y, z) - \int F(y, w)\  d\gamma(w) \right] d\pi(y, z) = 0, $$
which is implementable in terms of samples (see \cite{tabak2022distributional} for more details):
$$ \forall F(y, z) \in \mathcal{F}, \ \sum_{i} \left[F\left(y_i, z_i\right) - \frac{1}{N}\sum_k F\left(y_i, w_k\right) \right] = 0, $$
where $\mathcal{F}$ is a class of functions adapted to the number and distribution of the samples available. 

\item Same as above but with just one special test function $F_0$:
$$ F_0(y, z) = \rho(y | z) $$
(See in \cite{tabak2022distributional} how the condition that the vanishing of the non-negative quantity
$$  \int \left[F_0(y, z) - \int F_0(y, w)\  d\gamma(w) \right] d\pi(y, z) $$
suffices to guarantee independence.)

\item The vanishing of the mutual information between $Y$ and $Z$:
$$ MI(Y, Z) = \int \log\left(\frac{\pi(y, z)}{\rho(y) \gamma(z)} \right) d\pi(y, z) = 0. $$

\end{enumerate}

The second and third options pose the independence condition in terms of the vanishing of a non-negative functional $\Theta(\pi)$, with two immediate advantages: on the one hand, it permits adopting a standard penalty optimization procedure \cite{nocedal1999numerical}, and solve
$$ \min_T C(x, T(x, z)) + \lambda \Theta(\pi_T), $$
where 
$$ \pi_T(y, z) = T\# \pi(x, z) $$
is the push forward by the map $T$ of the original joint distribution $\pi$, and $\lambda > 0$ is a penalization parameter.
On the other hand, it allows us to enforce a detailed notion of independence. When $z$ has more than one component, the independence between $y$ and $z$ implies that $y$ should be independent of any subset of the $\{z^l\}$, including each individual factor alone. Because of the non-negative nature of the functional $\Theta$, we can enforce each of these requirement separately, writing
$$ \min_T C(x, T(x, z)) + \sum_k \lambda_k \Theta_k(\pi_T). $$
In option 3, for instance, $\Theta_k$ measures the mutual information between $y$ and the $k$th subset of the $\{z^l\}$ considered.

This is important for the barycenter problem in general, and more so for the hierarchical one. When the number of factors $\{z^l\}$ is large and the number of observations is comparatively small, most $\{z_i\}$ will be far from each other. This makes a global characterization of independence necessarily inaccurate, while assessing the independence between $y$ and lower dimensional subsets of the $\{z^l\}$ may be within reach. This statement applies even more strongly to the hierarchical barycenter problem, where some of the $\{z^l\}$ are available for only a fraction of the data.

Option 1 addresses the same issue automatically, since the set of all measurable functions $F(y, z)$ includes those that depend only on any given subset of the $\{z^l\}$. Hence all the options above are suitable candidates, each with its own advantages and challenges. For concreteness, we restrict attention to option 3, since this article's main objective is to develop the hierarchical barycenter's conceptual framework, rather than perfecting one technical approach or another.

Thus we will pose the problem in the form
\begin{equation}\label{eq:BarMI}
 \min_T L[T]= C(X, Y) + \sum_k \lambda_k MI\left(Y, Z_k \right), \quad Y = T(X, Z).
\end{equation}
Here the $\{Z_k\}$ represent subsets of the full set of cofactors $\{Z^l\}$.
Their choice, as well as the values of the penalization parameters $\{\lambda_k\}$ and the data-based formulation of the problem are discussed below.

\section{Related work}

The methodology presented here has partial overlap with two well established procedures: multi-task learning (MTL) \cite{zhang2021survey,caruana1997multitask} and transfer learning (TL) \cite{zhuang2020comprehensive,weiss2016survey}. Both methodologies  aim to exploit common information shared by different data sets and both are usually framed in terms of learning tasks. The main difference between the two is that while MTL aims to learn multiple tasks at the same time, TL leverages information available from an initial task to improve the performance of another, single target task. 

In this sense, the hierarchical barycenter (HB) has points in common with both procedures: it uses information from different data sets to enhance the estimate of the conditional density underlying one --any-- data set. This data set does not have to be known a-priori as in TL:  the information is ``transferred" by the pull back of the barycenter to one of the specific data sets (see eq. \ref{eq:prediction}). 

There are many differences between HB and the MTL-TL procedures. Starting with MTL, the problem is usually formulated through an objective function
\begin{equation}\label{eq:MTL}
MTL(w)=\sum_{k=1}^{l}L_{k}(f_{k}(w_{k})) + \lambda R(w),
\end{equation}
where $L_{k}$ is a loss function for the $k$-th task and the $w_{k}$ are the corresponding parameters for label's estimation. The function $R(w)$ regularizes the parameters within each task and specifies how much the parameters referring to different tasks are related. Different forms of MTL are represented by different functions $R$. The use of (\ref{eq:MTL})  assumes that there are features that are related to all tasks and that these features are learned by optimizing (\ref{eq:MTL}) (see \cite{zhang2021survey,argyriou2006multi}).

A first difference between this set up and the HB is that the latter does not impose a parametrization for each task. Since the pushforward is defined through the flow associated to the minimization in (\ref{Barycenter}), HB is non-parametric (see Section \ref{sec:Alg} for more details). From this point of view, HB is more similar to \cite{mejjati2018multi}, where task-specific estimators are considered as random variables and the task relationships are discovered by measuring the statistical dependence between each pair of random variables. The overall goal is to leverage information between random variables displaying higher degrees of dependence. As in our approach, the work in \cite{mejjati2018multi} uses kernel approximation of the mutual information \cite{kraskov2004estimating,belghazi2018mutual}, yet the scope is different. In HB, the mutual information is used to enforce the pushforward condition relating the marginals to the barycenter rather than to measure the degree of dependence.  

Another major difference between the HB and MTL is the barycenter itself. A byproduct of HB, absent in both MTL and TL, is the merging of multiple datasets into the barycenter. This distribution is characterized in a precise way, representing all the variability that cannot be explained by any of the cofactors across the different datasets. The barycenter represents a tool for at least two important tasks: the removal of variability \cite{TT2} and factor discovery \cite{yang2022conditional}. Section \ref{sec:HiddenFactrs} shows examples in which we use HB to remove the variability explainable by known cofactors to reveal a hidden signal in the data. The work in \cite{yang2022conditional} shows then how one can look for hidden factors that explain the hidden signal.  

We do not claim that HB necessarily leads to better results than MTL in scenarios in which both procedures apply. We propose the hierarchical barycenter as a general, first principled procedure, based on the mathematical theory of optimal transport, well-suited to the analysis of heterogeneous datasets.

\section{The Algorithm}\label{sec:Alg}

\subsection{Problem formulation in terms of samples}

In order to develop a data-driven methodology to solve the hierarchical barycenter, we need to transform the formulation in (\ref{eq:BarMI}) into one that uses not the joint distribution $\pi(X, Z)$ itself but samples thereof $(x_i, z_i), \ i \in \{1, \ldots. N\}$ and their image,  $y_i = T(x_i, z_i)$, under the unknown map. This transformation is straightforward for the transportation cost:
$$ C(X, Y) = \int c\left(x, T(x, z)\right) \pi(x, z) \ dx dz \rightarrow \frac{1}{N} \sum_i c(x_i, y_i). $$

Similarly, for the mutual information $MI(Y, Z^k)$, where $Z^k$ is a subset of the variables $Z$, we can write
\begin{multline}
MI(Y, Z^k) = KL\left[ \pi^k_T\left(Y , Z^k\right) || \mu\left(Y\right) \nu\left(Z^k\right)\right]  \approx \\ \approx \frac{1}{N_k} \sum_{i \in I_k} \Bigg[  \log\left[ \pi^k_T\left(y_i, z^k_i\right)\right] - \log \left[\mu\left(y_i\right) \nu\left(z^k_i\right)\right] \Bigg],
\end{multline}
where $I_k$ represents the subset of observations where all covariates $z^l \in Z_k$ are defined, and $N_k = |I_k|$, their cardinality. 
For this formulation to depend only on the data points, we replace the probability densities by their kernel-based estimation:
\begin{equation}\label{eq:KernelMI}
MI^{est}\left(Y, Z^k\right)  =  \frac{1}{N_k} \sum_{i \in I_k} R(y_{i},z_{i}^{k})  
\end{equation}
where 
\begin{multline}\label{eq:R}
R(y_{i},z_{i}^{k}) = \log  \left( \frac{1}{N_k} \sum_{j \in I_k} K^y\left(y_i, y_j\right) K^z\left(z^k_i, z^k_j\right) \right) \\
 - \log \left( \frac{1}{N_k} \sum_{j \in I_k} K^y\left(y_i, y_j\right) \right) - \log \left( \frac{1}{N_k} \sum_{j \in I_k} K^z\left(z^k_i, z^k_j\right) \right).
\end{multline}
Then problem (\ref{eq:BarMI}) adopts the data-driven form
\begin{equation}\label{eq:BarMI_sampl}
 \min_{\{y_i\}} L = \frac{1}{N} \sum_i c(x_i, y_i) + \sum_k \lambda_k  \frac{1}{N_k} \sum_{i \in I_k} R(y_{i},z_{i}^{k}) . 
\end{equation}

Notice that we do not need to keep the third term in $R(y_{i},z_{i}^{k})$, since it depends only on $z$ and we are minimizing over $y$. Without the third term, we are minimizing the log-likelihood of $\rho(z|y)$ over the variable $y$ on which we are conditioning. In other words, we are looking for the $y$ such that the observed $z$, given $y$, is least likely. It follows from the argument that this $y$ must be independent of $z$.

\smallskip  

In order to guarantee independence between $Y$ and $Z$, the choice of the covariate subsets $\{Z_k\}$ must satisfy the requirement that, for all subsets of the observations $\{x_i\}$, their maximal common subset of covariates must be one of the $\{Z_k\}$. This is just a minimal requirement though: one can add additional subsets --all the way to those consisting of individual $z^l$'s-- in order to enforce a more detailed notion of independence.
On the other hand, even the minimal requirement can sometimes be computationally unfeasible. For instance, when  covariate values are missing at random, we may be forced to consider all subsets of the $\{Z^l\}$, which grows exponentially with the number of covariates, and which may include subsets with corresponding sample sets $I_k$ with only a handful of available samples. In this case, a relaxation of the problem is appropriate, such as including only those subsets $\{Z_k\}$ with cardinality smaller than a prescribed small number $L_{max}$ and where the number of samples $|I_k|$ is larger than a minimum value $N_{min}$. The first condition disregards complex dependence between $Y$ and $Z$ involving the non-additive interaction of more than $L_{max}$ factors. Setting $L_{max} = 1$, for instance, would relax the notion of independence between $Y$ and $Z$ to that of independence between $Y$ and each individual $Z^l$. The second condition addresses possible over-fitting of small subsets of the data.

\subsection{Minimization through regularized gradient descent}\label{sec:minimization}

We minimize the objective function $L$ in (\ref{eq:BarMI_sampl}) through gradient descent, accelerated by a preconditioning procedure that can be conceptualized as a simplified version of implicit gradient descent: 
\begin{equation}
y = y - \eta \left( I + \eta H_d \right)^{-1} G,
\label{IGD}
\end{equation}
where $G$ with $G_i = \frac{\partial L}{\partial y_i}$ is the gradient of $L$ and $H_d$ is a diagonal matrix containing only the diagonal elements ${H}_i^i = \frac{\partial^2 L}{\partial y_i^2}$ of the Hessian matrix $H$. Equation (\ref{IGD}) with the full matrix $H$ instead of the diagonal $H_d$ is the building block of implicit gradient descent (see \cite{essid2023implicit} for a similar procedure for minimax problems and \cite{hanzely2022damped, MoreJorge1978} for the closely related Levenberg-Marquardt regularization of Newton’s method). Keeping just the diagonal elements of $H$ eliminates the need to invert an $n \times n$ matrix, while preserving to a large degree the regularizing effect of the implicit procedure.

The learning rate $\eta$ in (\ref{IGD}) is chosen adaptively according to the strategy described in \cite{essid2023implicit}. Far from the local extreme of the loss function, $\eta$ is small and (\ref{IGD}) reduces to regular gradient descent. Near the minimum, $\eta$ increases, converging to a (quasi) Newton method with faster convergence rate.  

The use of (\ref{IGD}) for our problem is greatly facilitated by the fact that the gradient of $L$ in (\ref{eq:BarMI_sampl}) can be robustly approximated by
\begin{equation}
 \frac{\partial L}{\partial y_i} = \frac{1}{N} \frac{\partial}{\partial y_i} c(x_i, y_i) + 
  \sum_{k / i \in I_k}   \frac{\lambda_k}{N_k}  \frac{\partial}{\partial y} \left[\log \left(\frac{\sum_{j \in I_k} K^y\left(y, y_j\right) K^z\left(z^k_i, z^k_j\right)}{\sum_{j \in I_k} K^y\left(y, y_j\right)} \right) \right]_{y = y_i}.
  \label{gradL}
\end{equation}
The simple form of the gradient of $L$ is due to the fact that the derivative of the mutual information with respect to the second argument of the kernels, namely with respect to the position of the centers, is a random variable with zero mean and vanishing variance as the number of samples grows.
We provide here a general argument of why this is the case; a more detailed derivation can be found the appendix. 

We focus on the second term in (\ref{eq:R}), since the result for the first term follows from exactly the same logic. Suppose that the first and second argument of $K^y$ are not necessarily computed at the same set of points $\{y_i\}$. This amount to substituting the estimation of $\mu(y)$, relative to the random variable $Y$, with kernels that are centered around a set of points $\{y'_{j}\}$, whose distribution $\xi(y)$ is different from $\mu(y)$. The second term in (\ref{eq:R}) can then be interpreted  as the relative entropy between $\mu$ and $\xi$:
\begin{equation}\label{eq:re}
\frac{1}{N_k}\sum_{i\in I_{k}} \log \left( \frac{1}{N'} \sum_{j=1}^{N'}K^{y}(y_i, y_{j}) \right)\approx \int \mu(y)\log(\xi(y))dy.
\end{equation}
Taking the derivative with respect to the second argument of $K^{y}$, the position of the centers, can therefore be thought, in the limit of $N' \rightarrow \infty$, as computing the variational derivative with respect to $\xi$ of the RHS of (\ref{eq:re}). Since the relative entropy is maximized when $\xi = \mu$ almost everywhere, then we have that
$$
\frac{\delta}{\delta \xi} \left.\int \mu(y)\log(\xi(y))dy\right|_{\xi=\mu} = 0.
$$
(The true relative entropy has an additional term involving the entropy of $\mu$, but this is immaterial for differentiation with respect to $\xi$.)

This justifies the approximation in (\ref{gradL}),
\begin{eqnarray*}
\frac{\partial L}{\partial y_i} &=& \frac{1}{N}\frac{\partial}{\partial y_i} c(x_i,y_i)  \\
&& +  \sum_{k / i \in I_k}   \frac{\lambda_k}{N_k}  \left[\frac{1}{\sum_{j\in I_k} K^y(y,y_j) K^z(z_i^k,z_j^k)}  \frac{\partial}{\partial y}  \left( K^y\left(y, y_j\right) K^z\left(z^k_i, z^k_j\right) \right) \right]_{y = y_i}\\
&& -  \frac{\lambda_k}{N_k} \left[ \frac{1}{\sum_{j\in I_k} K^y(y,y_j) }  \frac{\partial}{\partial y} K^y\left(y, y_j\right)   \right]_{y = y_i}.
\\
\end{eqnarray*}
which neglects to consider the derivative of the objective function with respect to the Kernel's second argument.

Similarly,  the diagonal second order derivatives in $H_d^i$ are well-approximated by
\begin{eqnarray*}
\frac{\partial^2 L}{\partial y_i^2} &=& \frac{1}{N}\frac{\partial^2}{\partial y_i^2} c(x_i,y_i)  \\
&& +  \sum_{k / i \in I_k}   \frac{\lambda_k}{N_k}  \left[\frac{\partial }{\partial y}\left( \frac{1}{\sum_{j\in I_k} K^y(y,y_j) K^z(z_i^k,z_j^k)}  \frac{\partial}{\partial y}  \left( K^y\left(y, y_j\right) K^z\left(z^k_i, z^k_j\right) \right) \right)\right]_{y = y_i}\\
&& -  \sum_{k / i \in I_k}   \frac{\lambda_k}{N_k} \left[ \frac{\partial }{\partial y}  \left( \frac{1}{\sum_{j\in I_k} K^y(y,y_j) }  \frac{\partial}{\partial y} K^y\left(y, y_j\right)   \right)\right]_{y = y_i} .
\end{eqnarray*}

\subsection{Prediction through map inversion}

The solution of the barycenter problem consists of two elements: the barycenter $\mu(y)$ itself
and the map $T(x; z)$ pushing forward the conditional distribution $\rho(x|z)$ to $\mu$. In our numerical solution, neither $\mu$ nor $T$ are given in closed form. Instead, they are both represented by the set $\{y_i\}$, where each $y_i$ is both an independent sample from the barycenter $\mu(y)$ and the value that the map $T$ adopts when applied to the sample pair $(x_i, z_i)$.

The barycenter has much value, as elaborated through an example in Section \ref{sec:HiddenFactrs}. Yet our original goal was not to explain variability away from $x$, but to simulate  $\rho(x|z^{*})$ for a given target value $z^{*}$. For this, we need to compute the inverse map $T^{-1}(\cdot,z^{*})$ and use it to push back all the points $\{y_{i}\}$ in the barycenter to obtain $N=\sum_{p}N_{p}$ samples $x_{i}(z^{*})$  from $\rho(x | z^{*})$.
It would appear at first that inverting a map $y=T(x, z)$ known only through samples $(x_i, z_i, y_i)$ should require numerical interpolation, e.g. using near-neighbors, kernel regression or neural networks. Yet it turns out that the structure of the minimization problem giving rise to $T$ provides a simple closed-form solution for $x = T^{-1}(y, z)$, obtained from the first order condition $\nabla_{Y}L=0$ of the objective function in (\ref{eq:BarMI}).

The condition that $\frac{\partial L}{\partial y_i} = 0$ in (\ref{gradL}) adopts the form
$$
 \frac{\partial}{\partial y} c(x_i, y) \Bigg|_{y = y_i} = 
- \frac{\partial}{\partial y} \sum_{k / i \in I_k}   \frac{\lambda_k N}{N_k} \log \left(\frac{\sum_{j \in I_k} K^y\left(y, y_j\right) K^z\left(z^k, z^k_j\right)}{\sum_{j \in I_k} K^y\left(y, y_j\right)} \right) \Bigg|_{\stackrel{y = y_i}{z = z_i}}.
$$
Notice that only the left-had side of this identity depends on $x_i$. It follows that, if $\frac{\partial c(x, y)}{\partial y} = a$ is invertible as $x = f(y, a)$,
then it provides a closed expression for $x$ in terms of $y$ and $z$, strictly valid for every sample pair $(y_i, z_i)$. Since this expression is smooth as a function of $y$ and $z$, it provides a natural, closed form expression for $x = T^{-1}(y, z)$ for pairs $(y, z)$ not in the sample set. In particular, under the canonical cost $c(x, y)=\|x-y\|^{2}/2$, we obtain
\begin{equation}
x = T^{-1}(y, z) = y + \frac{\partial}{\partial y} \sum_{k / i \in I_k} \frac{\lambda_k N}{N_k} \log \left(\frac{\sum_{j \in I_k} K^y\left(y, y_j\right) K^z\left(z^k, z^k_j\right)}{\sum_{j \in I_k} K^y\left(y, y_j\right)} \right).
\label{inversion}
\end{equation}

\subsection{Hyper-parameter tuning}\label{sec:HPtune}

One needs to choose a set of hyper-parameters for the data-driven formulation: bandwidths of kernels in $z^k$ and $y$ spaces  and penalty coefficients $\{\lambda_k\}$.  The following two subsections discuss their choice in the current framework.  

\subsubsection{Bandwidths in $z^k$-space}
These hyper-parameters must be chosen first, as their values remain constants throughout the procedure, since the $z_i$ are kept fixed during the iteration pushing forward $x$ to  $T(x, z)$.

We chose the bandwidths $h_{z}$ in $z$ space  based on Silverman's rule-of-thumb \cite{silverman2018density, duong2005cross} and the cardinality of the subset of the the data for which each specific component of $z$ is known. Then the bandwidth for $K^{z}(z_{i}^{k},z_{j}^{k})$ with $i, j\in I_{k}$, is given by a diagonal covariance matrix with elements $d_{kk}$ given by
\begin{equation}\label{eq:zbandwidth}
	\sqrt{d_{kk}} = \sigma_{z^k}\left(\frac{4}{d+2}\right)^{} |I_{k}|^{\frac{-1}{d+4}},
\end{equation}
where $\sigma_{z^k}$ is the estimated standard deviation over the $\{z_{i}^k:  i\in I_{k}\}$, $|I_{k}|$ is the cardinality of $I_{k}$ and $d$ is the dimension of $z$.

\subsubsection{Penalty coefficients $\{\lambda_{k}\}$ and bandwidths in $y$ space}
These two sets of hyper parameters are related, since both control the extent to which we resolve the independence of $y$ and $z$: the magnitude of $\lambda_{k}>0$ quantifies the weight assigned to the vanishing of the mutual information $MI(Y, z^{k})$, and the bandwidths set the accuracy of the estimation of $MI(Y, z^{k})$ via the kernels $K^{y}$ in (\ref{eq:KernelMI}). 

We tune $\lambda_{k}$ and $h_{y}$ via cross validation. To this end, we set aside from the data a validation set $\{x_{i,val}, z_{i,val}\}$, and use as objective function for the cross validation the log-likelihood of this set:
\begin{equation}\label{eq:CVobj}
\overline{\log \mathcal{L} }= \frac{1}{N_{val}} \sum_{i=1}^{N_{val}}\log \hat \rho(x_{i,val}| z_{i,val}),
\end{equation}
where for each choice of the hyper-parameters $\lambda$ and $h_y$, $\hat \rho(\cdot |z_{i,val})$ is estimated in following way: we solve the barycenter problem using the training data, obtain samples $ x_{i, val}^j \sim \rho(\cdot |z_{i,val})$ through the inversion of $T$ (described in the following subsection) on each $y_j$ in the barycenter and each $z_{i,val}$, and then estimate $\rho(\cdot |z_{i,val})$ through kernel density estimation on those $x_{i, val}^j$.

We adopt the following strategy in order to facilitate the exploration of a hyper parameter space that would otherwise be too high dimensional when $k$ takes more than 2 different values. All $\lambda_{k}$ are determined by a single tunable parameter $\lambda$, through
\begin{equation}\label{eq:lambdak}
\lambda_{k}=\lambda r_{i}\;\; \mbox{where} \;\; r_{i}=|I_{k}|MI^{est}(X,Z^{k}).
\end{equation}
The rationale for this choice is that the penalty coefficient should increase with the degree of dependence between $X$ and $Z^k$, quantified by their mutual information, and to the significance of the corresponding subset $I_k$, which its naturally quantified by its cardinality $|I_{k}|$.

Additional information regarding hyperparameter tuning will be provided when discussing  concrete examples in Section \ref{sec:Num}. 

\section{Numerical Examples}\label{sec:Num}
This section illustrates the use of the HB on two common classes of problems in the machine learning literature: prediction on data sets with missing values and on data sets with structured cofactors. In both cases, the output of our analysis are samples $\{x^*_i\}$ drawn from the conditional density $\rho(x|z^{*})$, obtained through the procedure described in Section \ref{sec:Alg}.

\subsection{Missing Data - Synthetic Example}\label{sec:MDSE}
Our first synthetic example uses a synthetic dataset $\{(x_{i},z_{i})\}_{i=1,..., N}$, where $x_{i}\in \R$ is sampled from a normal distribution $\mathcal{N}(f(z_{i}),g(z_{i})^{2})$ with mean $f(z)$ and standard deviation $g(z)$, and $z_{i}=(z_{i}^{1},z_{i}^{2})\in\R^{2}$ is drawn from a bivariate uniform distribution $U_{[0,1]^{2}}$ with independent components.  The dataset is then divided into three subgroups: 
\begin{itemize}
\item $I_1$, with $x_{i}$ and only $z_{i}^{1}$ observed,
\item $I_2$, with $x_{i}$ and only $z_{i}^{2}$ observed, and
\item $I_3$, with $x_{i}$ and both $z_{i}^{1}$ and $z_{i}^{2}$ observed.
\end{itemize}
The subsets $I_1$ and $I_2$ have 80 points each, while $I_3$ contains only 20 points. We also keep an additional set of 20 points for cross validation, as described in section \ref{sec:HPtune}, with the values of all variables known.

In this case, the objective function (\ref{eq:BarMI}) has the form
\begin{equation}
\min_{y_i} \sum_{i=1}^{N} c(x_i, y_i)  + \lambda_1 MI^{est}(y,z^1) + \lambda_2 MI^{est}(y,z^2) + \lambda_3 MI^{est}(y,(z^1,z^2)) + \lambda_0 MI^{est} (y,w),
\end{equation}
where $w_i$ is a categorical variable indicating whether data point $x_{i}$ belongs to either $I_{1}$, $I_{2}$ or $I_{3}$. As described in Section \ref{sec:extension}, including $w$ precludes the resulting barycenter from becoming a mixture of three distinct distributions, since the variability due to $w$ must be explained away. 

We run experiments with three different choices for the functions $f(z)$ and $g(z)$:
\begin{itemize}
\item Test 1 - Additive mean: $f(z) = 4z^1(1-z^1) + 0.5z^2,\quad
g(z) = 0.2.$
\item Test 2 - Non-additive mean: $
f(z) = 4z^1(1-z^1) + 0.5 z^1z^2,\quad g(z) = 0.2.
$
\item Test 3 - Heterogeneous standard deviation: $
f(z) = 4z^1(1-z^1) + 0.5z^2,\quad g(z) = 0.25(\sqrt{z^1}+\sqrt{z^2}).
$
\end{itemize}
The idea behind these tests is to quantify the procedure's accuracy under increasing complex levels of dependence between the random variables $X$ and $Z$. 

We compare our methodology with three alternative procedures for conditional density estimation, computing in all cases the Kullback-Leibler ($KL$) divergence (\cite{kullback1951information}) between the estimated and the exact density. Since the densities under consideration are Gaussian, we can use the following closed form for the $KL$ divergence for a given $z$ (see for instance \cite{robert1996intrinsic}): 
$$
KL(\hat \rho(\cdot |z ) || \rho(\cdot |z)) = \log\left(\frac{\sigma_\rho  }{\sigma_{\hat{\rho}}}\right)+ \frac{ \sigma_{\hat{\rho}}^2 + ( \mu_{\rho}-\mu_{\hat{\rho}})^2 }{2\sigma_{\rho}^2} - 1/2,
$$
where $\sigma_{\rho}$ and $\mu_{\rho}$ are the standard deviation and the expected values of $\rho$ respectively (similarly for $\hat{\rho}$). We then average $KL(\hat \rho(\cdot |z ) || \rho(\cdot |z))$ over different values of $z$ chosen on a uniform grid covering the support of the density of underlying $z$. The procedures under comparison are: 
\begin{enumerate}
\item Hierarchical Barycenter (HB), the procedure described in Section \ref{sec:Alg}.
\item  Benchmark 1 (B1), a regular barycenter problem which uses only the points in $I_{3}$ for which the values of both covariates $(z^{1},z^{2})$ are known.
\item  Benchmark 2 (B2), which first imputes the missing values for the covariates $z$'s (based on the nearest neighbor in $x$ space) and then solves a regular barycenter problem with no missing values (see subsection \ref{sec:DB})
\item Benchmark 3 (B3), which solves the classical barycenter problem without hiding any of the values of either $z^{1}$ or $z^{2}$. This is the best possible scenario, since everything is known and the error in the estimate of the conditional density is exclusively due to the Monte Carlo approximations used to compute the Mutual Information that enforces the independence between $y$ and $z=(z^{1}, z^{2})$. 
\end{enumerate}

The results are summarized in the following table:
\begin{table}[h!bp]
\centering
\begin{tabular}{c|c c c|c}
Procedures & HB  & B1 & B2 & B3\\\hline
Test 1 &     \textbf{0.1997}   &  0.7366  &     0.2189 &  0.1605  \\
Test 2 &   \textbf{0.1890}   & 0.7014    &  0.2335  & 0.1597  \\
Test 3 &    \textbf{0.2746}   &  0.5963   &   0.2817 &  0.1998   \\\hline
\end{tabular}
\caption{KL divergence for the three synthetic tests, which estimate the conditional density using the hierarchical barycenter (HB) and three different benchmarks (B1-B3) described in subsection \ref{sec:MDSE}.}
\label{tab::synthetic_missing}
\end{table}

Table \ref{tab::synthetic_missing} shows that the lowest $KL$ values between the exact and the estimated densities are obtained for the hierarchical barycenter, achieving values close  to those obtained using B3, where no values of the cofactors $z$ were hidden. The estimate using B1 has the largest error, since it does not use the information contained in $I_2$ and $I_3$

\subsection{Missing Data - Bone Mineral Density}\label{sec:MDBones}
This section analyzes a data set with spinal bone mineral density measurements on 485 North American adolescents \cite{hastie2009elements}.
 
\begin{figure}[ht!]
\subfloat[Complete data set]{
\includegraphics[width=0.45\textwidth]{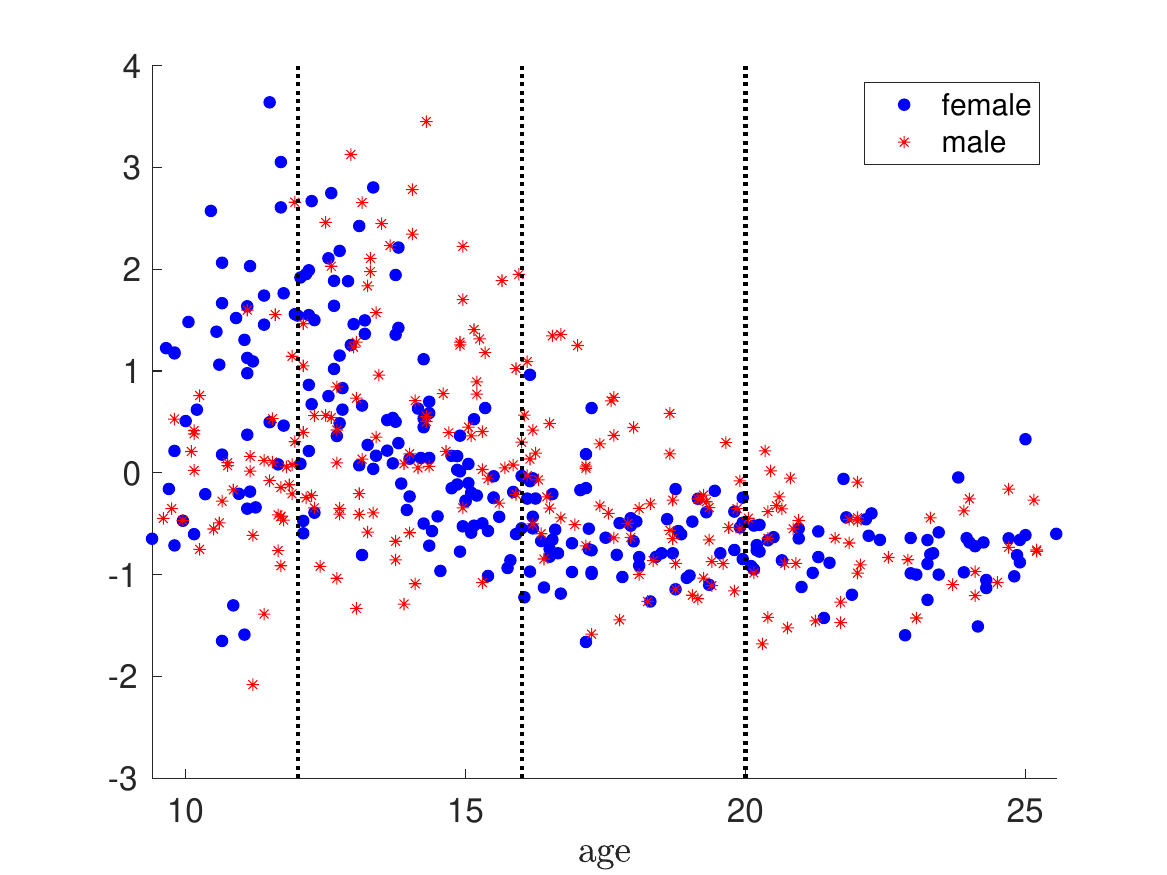}}
\subfloat[Data contained in $I_3$]{
\includegraphics[width=0.45\textwidth]{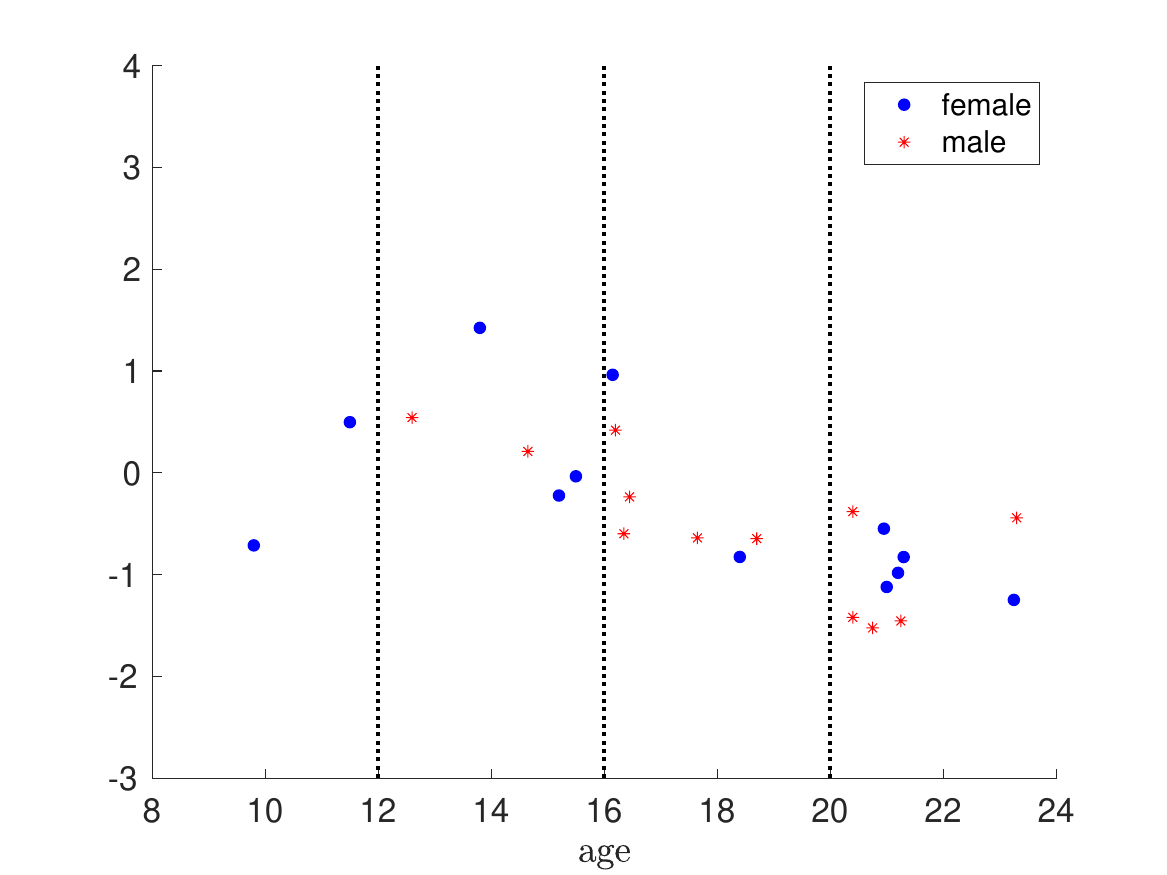}
}
\caption{Bone mineral density ($y$-axis) as a function of age ($x$-axis) and gender (color). The left panel displays the full data set --including those covariate values that will not be made available to the analyst--  while the right panel displays only the data in subset $I_3$, where the values of both $z^1$ and $z^2$ are known.}
\label{fig:bones}
\end{figure}

In order to adapt this data set to our purpose, we divide the data set into 3 subgroups and hide for each group the values of different cofactors, recreating the same setting of the synthetic example in subsection \ref{sec:MDSE}. In this context, $z^1$ and $z^2$ represent the gender and the age of each individual and $x$ represents the bone mineral density. The three subgroups  $I_{1,2,3}$ have the same meaning as in subsection \ref{sec:MDSE}. While $I_{1,2}$ both contain 218 points, the subgroup $I_3$, with no hidden values for $z^1$ or $z^2$, contains 24 points. In addition, we have a validation set of 25 points. Figure \ref{fig:bones} depicts the data set, highlighting the much smaller subset where all information is available. 

Since there is no ground truth to compare the results with, we assess the accuracy of the estimated conditional densities through the corresponding likelihood of the test set. In order to have more robust statistics, we repeat the experiment 30 times, each time hiding the values of the cofactors for a different subset of points. The results reported in the Table \ref{tab:bone} contain the average likelihood over those different realizations of $I_{1,2,3}$. Again, other than B3, which uses complete information supposedly not available,  the hierarchical barycenter yields the smallest error. 
\begin{table}[h!bp]
\centering
\begin{tabular}{c|ccc|c}
Approaches & HB  & B1 & B2 & B3\\\hline
 & \textbf{-1.0790}  &-1.3671  &   -1.2573& -0.8596\\\hline
\end{tabular}
\caption{Average likelihood of the validation set for the bone mineral dataset (The higher the better.) }
\label{tab:bone}
\end{table}

\begin{figure}[h!bp]
\centering
\subfloat[Results from HB]{
\includegraphics[width=0.7\textwidth]{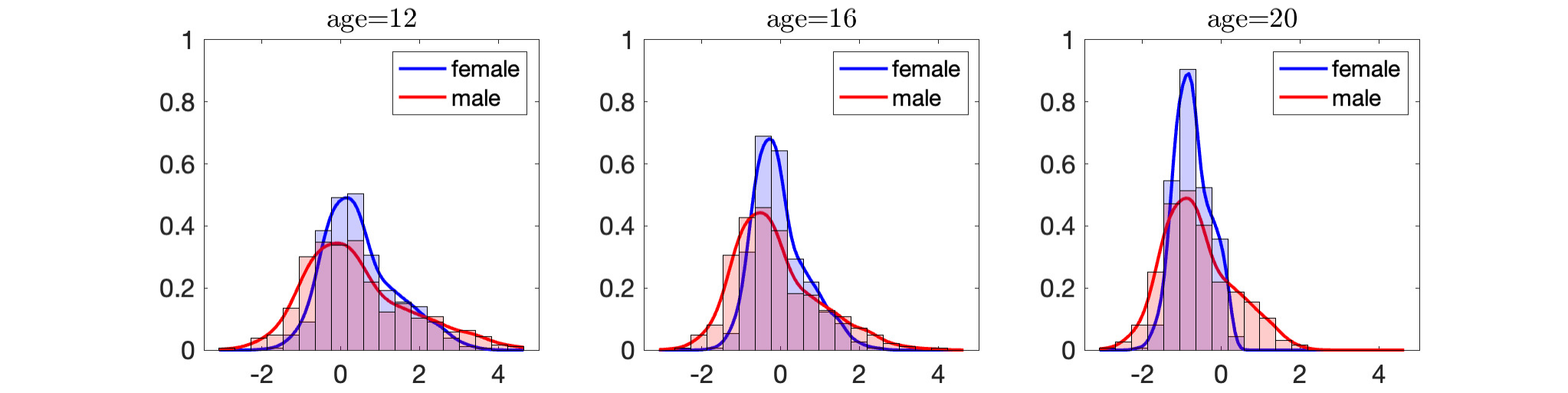}}\\
\subfloat[Results from B1]{
\includegraphics[width=0.7\textwidth]{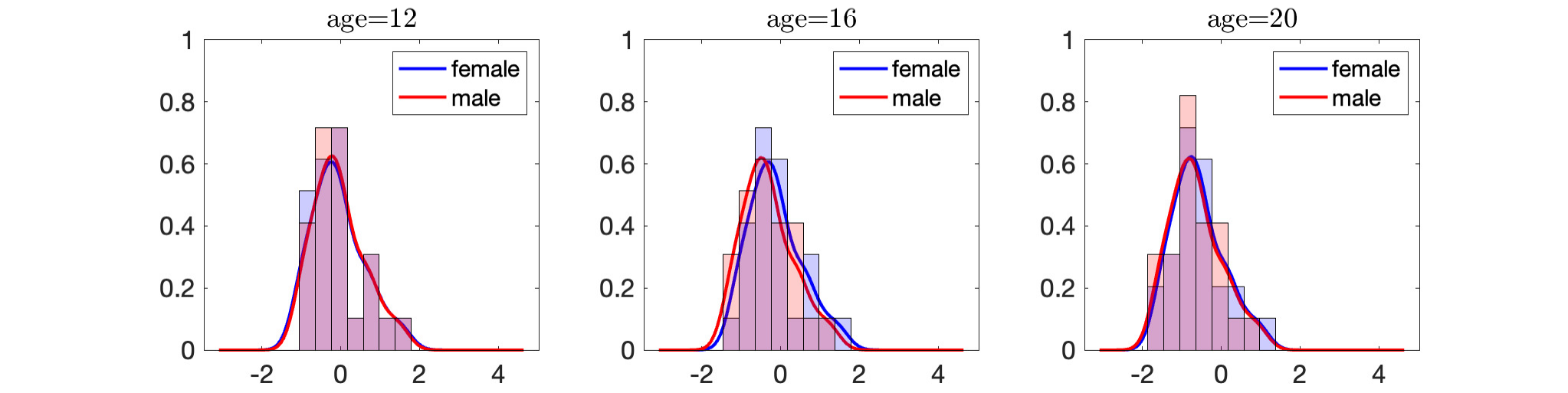}}\\
\subfloat[Results from B3]{
\includegraphics[width=0.7\textwidth]{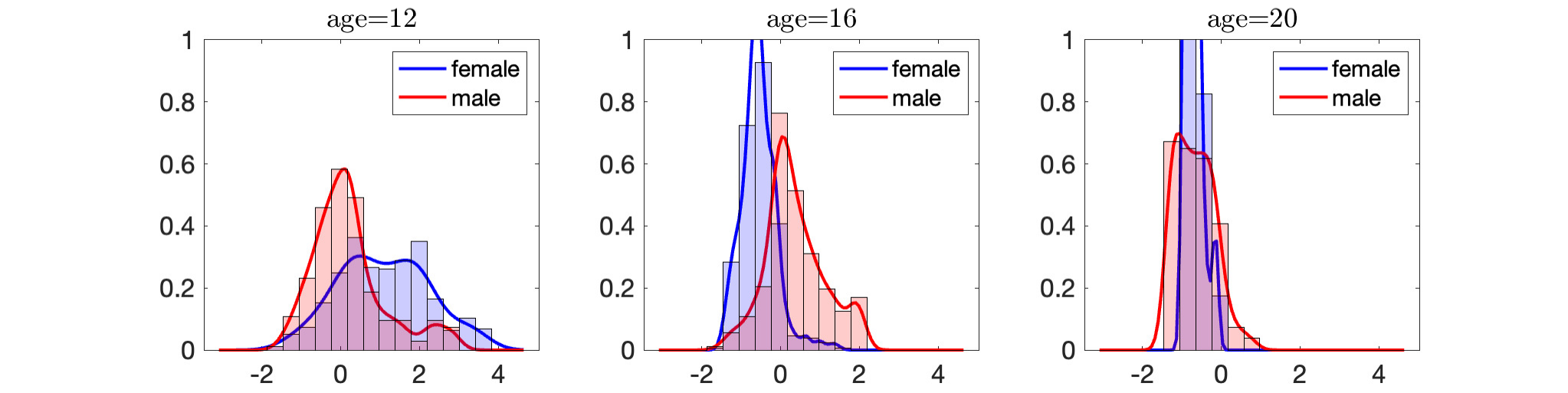}}
\caption{Conditional densities $\rho(x|z^{*})$ estimated for three different ages and both genders of bone mineral density dataset. The conditional densities are visualized through histograms and a kernel density estimator using the samples from $\rho(x|z^{*})$ obtained with the hierarchical barycenter and the benchmark procedures 1 and 3 respectively.}
\label{fig:result_bone}
\end{figure}

Figure \ref{fig:result_bone} visualizes some of the conditional densities obtained for three different ages, using the hierarchical barycenter and Benchmarks 1 and 3. 
Even though it is hard to asses visually from this particular instance whether HB is performing better than B1 when B3 is used as surrogate for the ground truth, one feature that HB seems to reproduce better than B1 is the heteroscedasticity of the conditional distribution for females, namely the decrease of its standard deviation as age increases. 

\subsection{Structured cofactors -  synthetic examples}\label{sec:SCSE}

\subsubsection{Different cofactors}\label{sec:diffcofactors}
\begin{figure}[h!bp]
\centering
\includegraphics[width=0.4\textwidth]{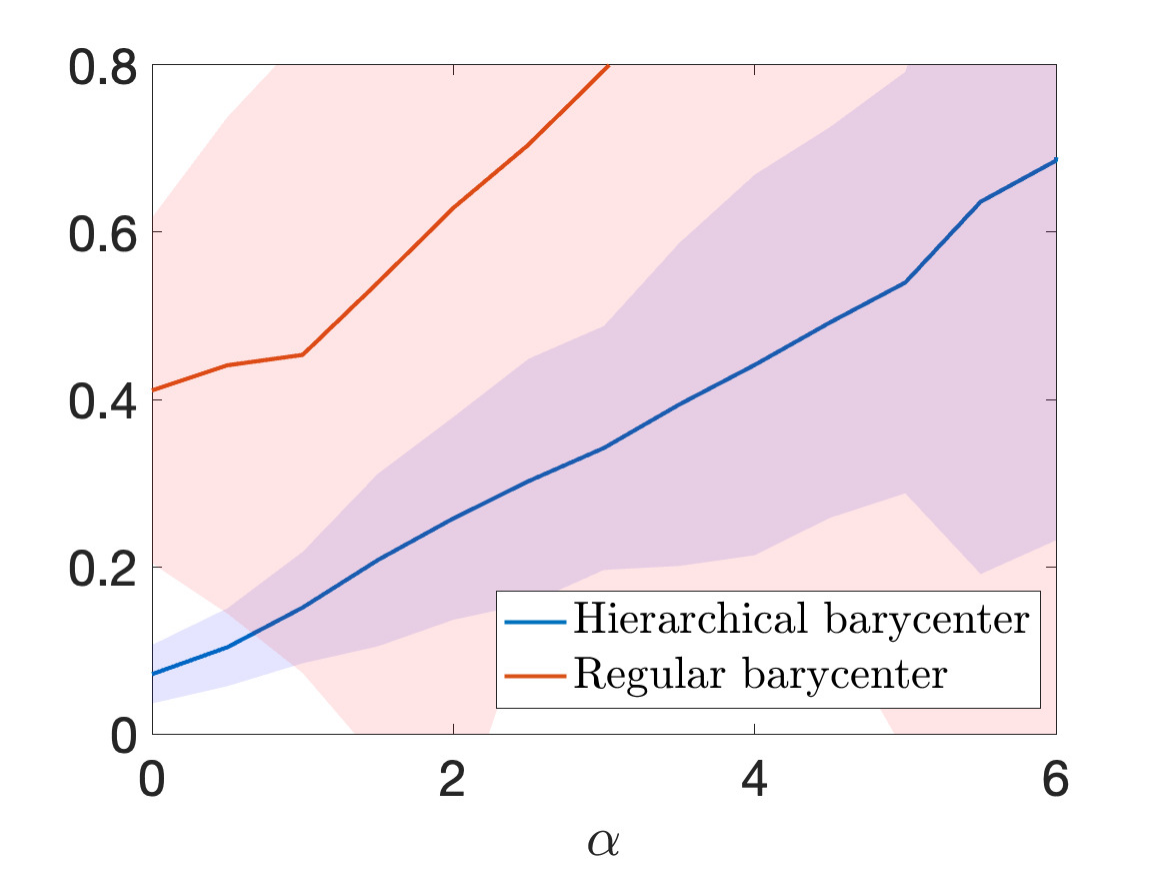}
\caption{The solid lines, relative to hierarchical barycenter and the regular barycenter respectively, represent the average KL value over different realizations of the noise. The shaded area corresponds to mean plus minus one standard deviation over 30 different realizations of the noise.}
\label{fig:KL_all}
\end{figure}

This section tests the hierarchical barycenter on datasets where different subsets have different cofactors. In particular, we consider a dataset divided into two subgroups:
\begin{itemize}
\item $I_{1}$, where $x_{i}$ is drawn from $\mathcal{N}(f_{1}(z),0.2^{2})$, with $z\in\R^{2}$ and $f_{1}(z)=4z^1(1-z^1) + \alpha (z^2-1/2)$
\item $I_{2}$, where $x_{i}$ is drawn from $\mathcal{N}(f_{2}(z^1),0.2^{2})$, with $z^1\in\R$ and $f_{2}(z^1)=4z^1(1-z^1)$
\end{itemize}
The variables $z^1$ and $z^2$ are drawn independently from $U_{[0,1]}$. The value of $\alpha \geq 0$ controls how much information is shared between the two subgroups. The goal is to estimate the conditional density of points in $I_{1}$ with aid from the information available in $I_{2}$.  For large values of $\alpha$, one would expect little gains from using the information contained in $I_{2}$ in order to estimate the conditional density of points in $I_{1}$, an effect balanced by the fact that the set $I_{1}$ contains only 40 sample points, while $I_{2}$ contains 100, yielding potential value to its use. Hyper-parameter tuning is performed via cross validation over the points in $I_{1}$. Figure \ref{fig:KL_all} displays the KL divergence between the true value of $\rho(x|z)$ and its estimation via the hierarchical barycenter using the points in both $I_{1}$ and $I_{2}$, and the classical barycenter using only the points in $I_{1}$. In order to mitigate the effect of specific realizations of the training set, we average the value of the KL over 30 realizations of the noise and over all the values of $z$ in $I_{1}$. As expected, the advantage of using the information contained in $I_{2}$ decreases as $\alpha$ increases.  

\subsubsection{Extrapolation}
\begin{figure}[h!]
\centering
\includegraphics[width=0.5\textwidth]{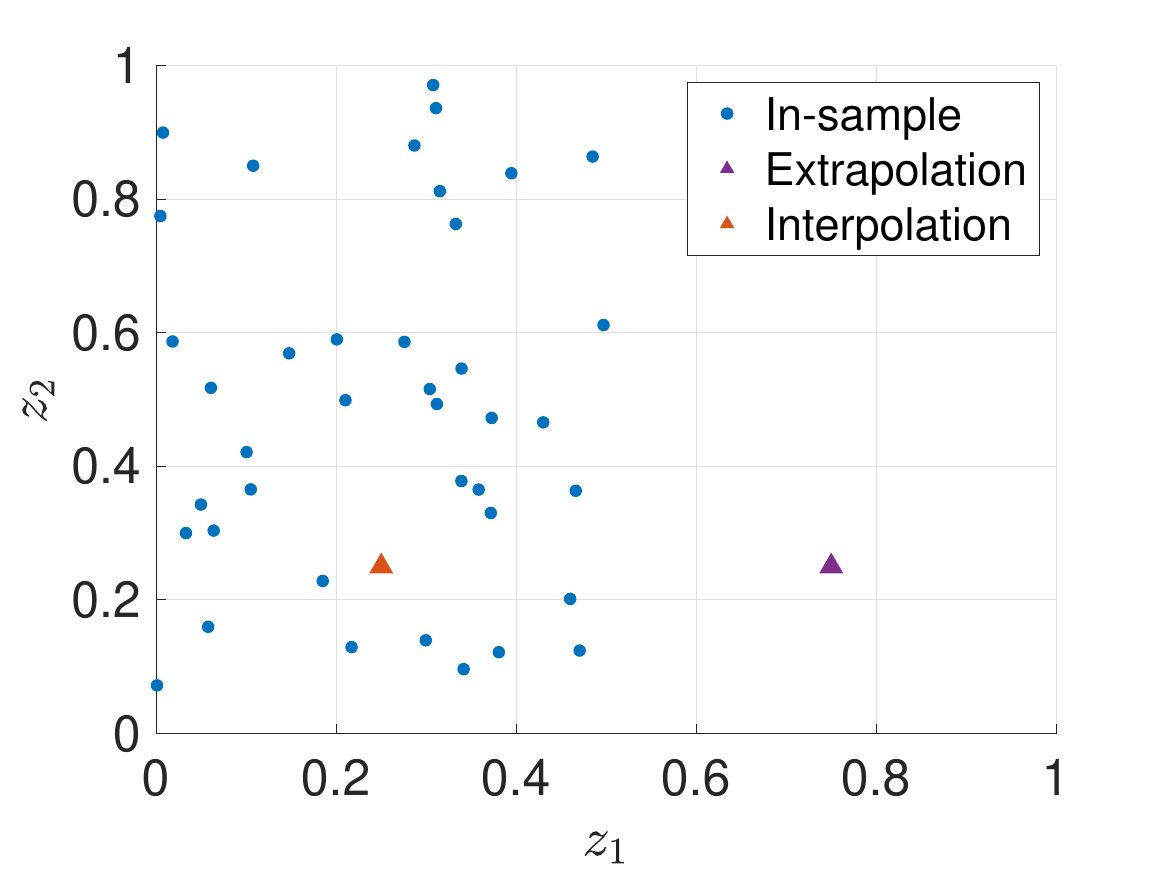}
\caption{Scatter plot of in-sample observations relative to $(z_1,z_2)$ in $I_1$. The triangular markers indicate the target values $z^{*}$ for which the estimation of $\rho(x|z^{*})$ is sought.}
\label{fig:distribution_sample}
\end{figure}

\begin{figure}[h!]
\centering
\subfloat[Hierarchical barycenter using points in $I_1\cup I_2$]{
\includegraphics[width=0.7\textwidth]{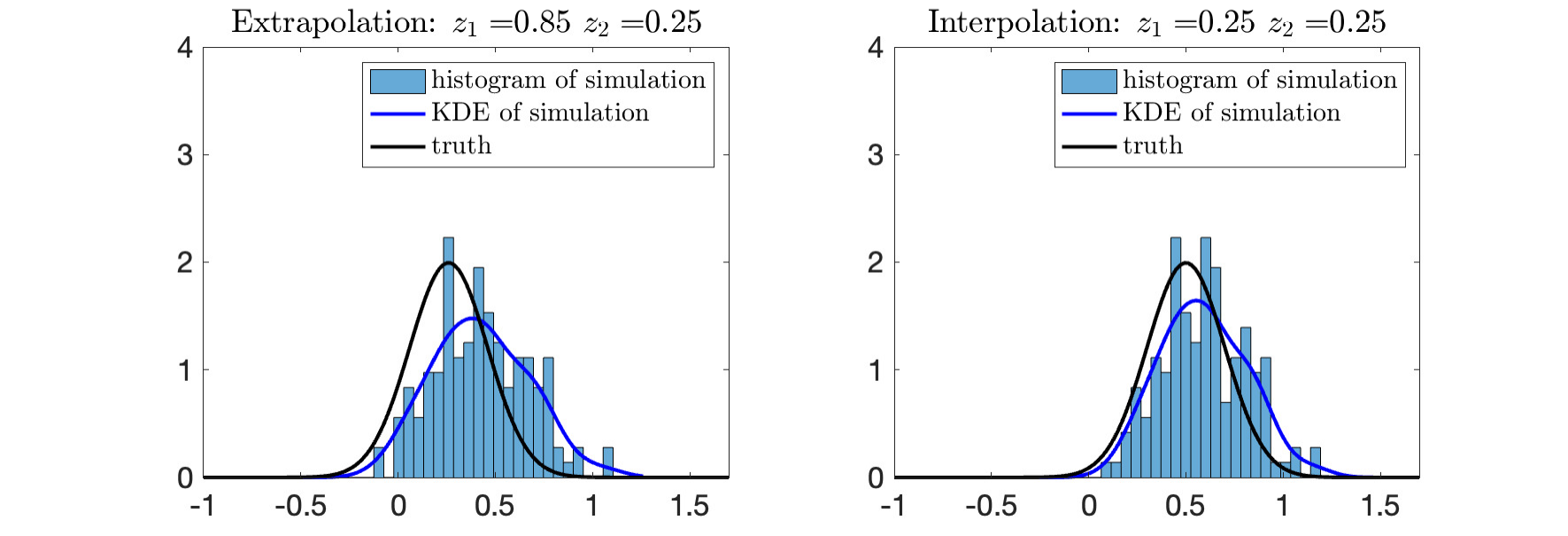}}\\
\subfloat[Regular barycenter using only points in $I_1$]{
\includegraphics[width=0.7\textwidth]{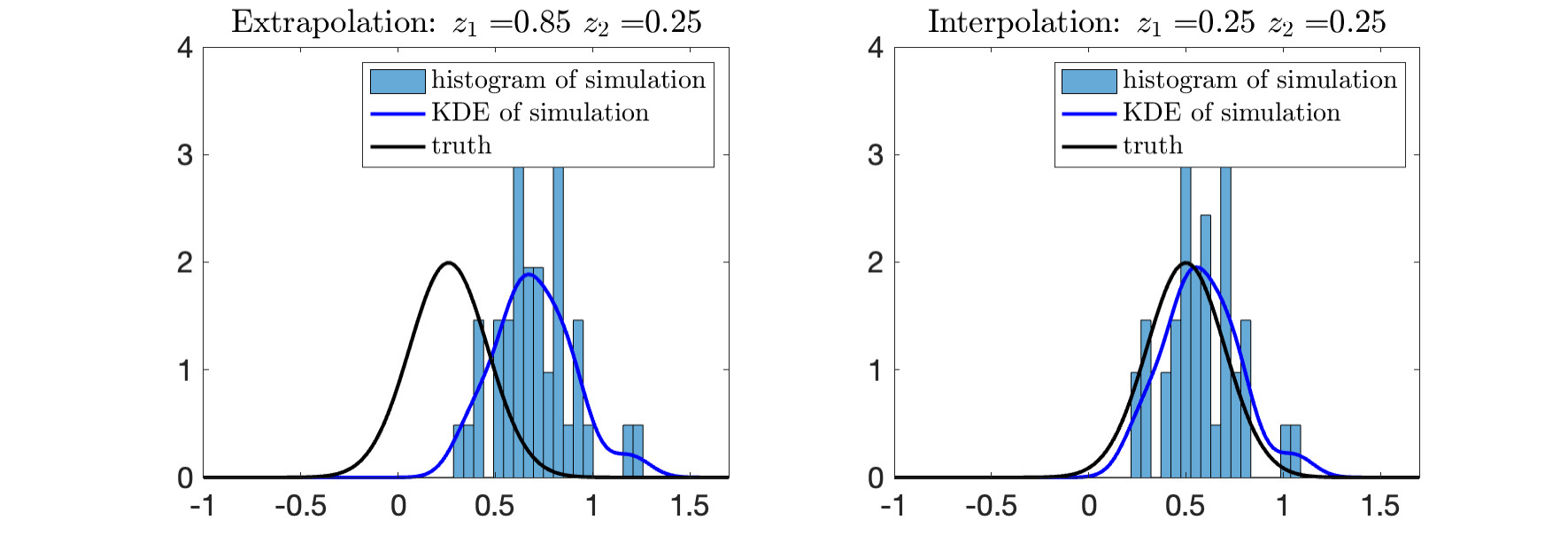}}
\caption{The histograms are relative to simulated points through the barycenter (a) and the hierarchical barycenter (b) respectively (see description of the procedure in Section \ref{sec:Formulation}). The blue curves are obtained performing kernel density estimation on the simulated points. The ground truth, known in closed form in this case, is represented by the black solid curve.}
\label{fig:extrapolation}
\end{figure}

We modify the experiment in sub-section \ref{sec:diffcofactors}, setting $\alpha=1$ and changing the distribution of the $z$ as follows: 
for the points in $I_1$, $z_1\sim U_{[0,0.5]}$ and $z_2 \sim U_{[0,1]}$ while for the points in $I_{2}$ we have $z_1\sim U_{[0,1]}$. The goal is to estimate $\rho(x|z_{*})$ underlying the points in $I_{1}$ when $z_{*}^{1}>0.5$ and $z_{*}^{2}\in[0,1]$ . The samples in $I_{1}$  are missing  information for such estimate, since $I_1$ does not contain any points with $z_1\in [0.5,1]$. Since such information is instead contained in $I_{2}$, one can hope to use the points in $I_{2}$ to extrapolate $\rho(x|z_{*})$. Figure \ref{fig:distribution_sample} displays the data set (blue points) together with the two points used for the interpolation and extrapolation. 
Figure \ref{fig:extrapolation} compares the estimate of $\rho(x|z_{*})$ underlying the model used to generate $I_{1}$ for $z_{a}=(0.85,0.25)$ and $z_{b}=(0.25,0.25)$. As expected, the estimate of  $\rho(x|z_{b})$ is close to the truth when using only the data in $I_{1}$ as opposed to the estimate of $\rho(x|z_{b})$ (see second column of the Figure \ref{fig:extrapolation}). When we use the solution of the regular barycenter problem to estimate $\rho(x|z_{a})$ computed only on the data relative $I_{1}$ the estimate is far from the truth (lower left panel). We need to extend the barycenter to the hierarchical barycenter to also use the points in $I_{2}$ (using the procedure described in Section \ref{sec:Alg}) in order to improve substantially the estimate of $\rho(x|z_{a})$ (upper left panel).
As in the previous section, the  hyper-parameter tuning is done through cross-validation on $I_1$. 

\subsection{Hierarchical barycenter and hidden factors}\label{sec:HiddenFactrs}
A significant byproduct of the hierarchical barycenter procedure is the barycenter itself. This is a distribution containing only the variability in $x$ that cannot be explained by the known cofactors $z$ (\cite{TT2}). Consider the following modification of the example described in Section \ref{sec:diffcofactors}:

\begin{itemize}
\item In $I_{1}$,  $x=4z^1(1-z^1) + (z^2-1/2) + 0.2 \epsilon$,
\item In $I_{2}$, $x=4z^1(1-z^1)+ 0.2 \epsilon$,  
\end{itemize}
with both random variables $z^{1}$ and $z^{2}$ uniform in $[0,1]$, and 
$$ \epsilon \sim N(z_{hidden},0.25^2) , \quad z_{hidden} \sim \frac{1}{3}\delta_{-1} + \frac{2}{3}\delta_1. $$

This example is very similar to the previous ones. The main  difference is that the noise involved is no longer Gaussian but it has  a bimodal distribution with two modes centered at $+1$ and $-1$ due to the $z_{hidden}$, a \emph{latent} random variable, i.e. one whose values are not measured.

We instead are given samples $(x_{i},z^{1}_{i},z^{2}_{i})$ and the goal is to see whether the bimodal pattern can be detected by looking at the barycenter or, in other terms, if we can characterize the variability of $x$ that is not due to the known $z^{1}$ and $z^{2}$. This goal can be achieved via the numerical  procedure developed in \cite{TT2, yang2022conditional} computing the -classical- barycenter of distributions $\rho(x|z^{1},z^{2})$ from samples contained in $I_{1}$. The question here is to see if we can integrate the information contained in $I_{2}$ via the hierachical barycenter. In the following, the numerical experiments are performed with $40$ points in $I_1$ and $50$ points in $I_2$.

\begin{figure}[h!bp]
\centering
\includegraphics[width=0.4\textwidth]{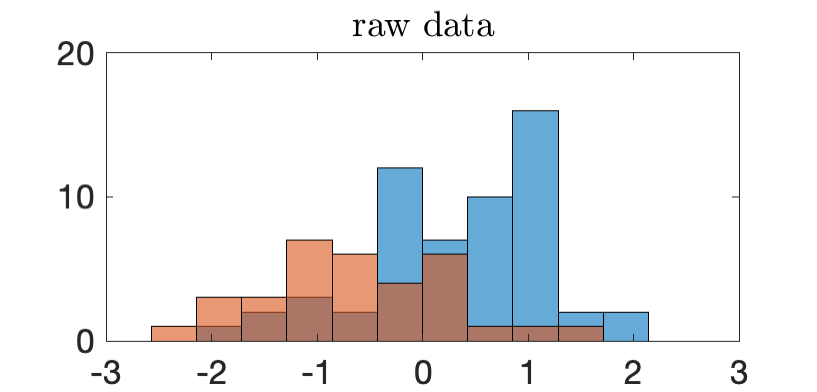}
\caption{Histogram relative to  $I_1\cup I_2$,  colored according to the hidden binary value.}
\label{Fig::bimodal_raw}
\end{figure}

Figure \ref{Fig::bimodal_raw} shows the histogram relative to the data set $I_1\cup I_2$. As expected the histogram does not look bimodal since the variability in $x$ derived from $z^{1}$ and $z^{2}$ hides the one due to $z_{hidden}$.

\begin{figure}[h!bp]
\centering
\includegraphics[width=0.8\textwidth]{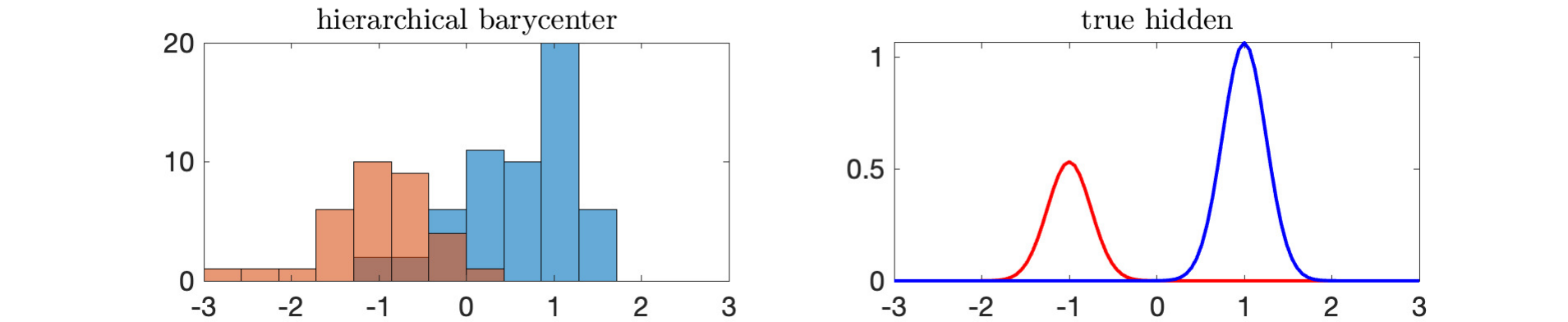}
\caption{Left: histogram of hierarchical barycenter computed using both $I_1$ and $I_2$,  after re-scaling the data.  Right: true distribution of $\epsilon$.  The two modes with $z_{hidden}=\pm 1$ are indicated via colors.  One can see how the hierarchical barycenter makes more evident the bimodality of the true distribution, which is hidden in the original data by the known covariates.}
\label{Fig::bimodal_hbary}
\end{figure}

Figure \ref{Fig::bimodal_hbary}  shows the barycenter obtained with the procedure described in Section \ref{sec:Alg}. 

\begin{figure}[h!bp]
\centering
\includegraphics[width=0.8\textwidth]{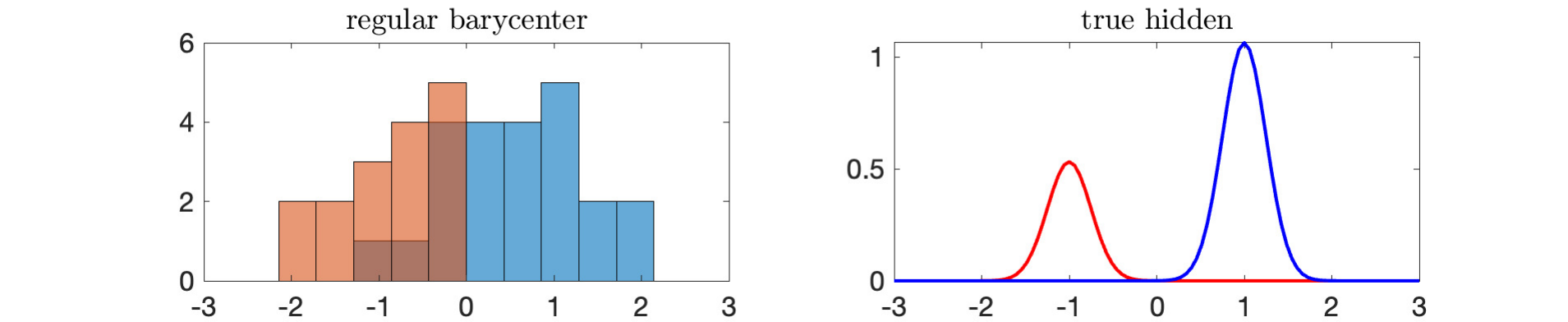}
\caption{Left: histogram of barycenter barycenter computed using only the samples in  $I_1$,  after taking z-score (just to normalize the scale).  Right: true distribution of $\epsilon$.  The two modes with $z_{hidden}=\pm 1$ are indicated via colors.  The regular barycenter does not show two modes very clearly because of the small sample size}
\label{Fig::bimodal_bary}
\end{figure}

We close this section with a numerical experiment comparing the ordinary barycenter computed without using the samples in $I_{2}$ and the hierachical barycenter that instead uses $I_{1}\cup I_{2}$. Figure \ref{Fig::bimodal_bary} displays the classical barycenter and, as it can be noticed, it is harder in this case to detect the bimodality of this distribution. This is due to the small size of $I_{1}$, showing that the information contained in $I_{2}$ in this case improved the detection of $z_{hidden}$.

\section{Conclusions}

The problem of inferring from data how a set of variables of interest $x$ depends on covariates $z$, is frequently formulated under the assumption that the observations consist of a set of identically distributed data pairs $\left\{z_i, x_i\right\}$. Yet the population of samples for real data is often strongly heterogeneous; in particular, the kind and number of the covariates $\{z_i\}$ may depend on the observation $i$. This arises for instance when data sets from various sources are aggregated, each with its own set of observed covariates, when some covariate values have not always been observed or recorded, and when the covariates have a hierarchical structure, so that only a subset of them is defined for each observation.
This article proposes and develops a methodology to address data analysis under such scenarios, simulating the conditional distribution $\rho(x|z)$ through an extension of the optimal transport barycenter problem to heterogeneous and not fully observed covariates $z$. 

Applying this methodology to a data set produces, in addition to a simulation of $\rho(x|z_*)$ for any target value $z_*$ --which may itself be incompletely observed-- samples $\{y_i\}$ from the barycenter $\mu$ of $\rho(:|z)$. The barycenter has additional applications, such as facilitating the detection and identification of hidden covariates. The corresponding variable $Y$ is defined as the one with minimal transportation cost from $X$ among all random variables independent of the covariates $Z$, where the latter includes additional markers of missing data.
The numerical procedure developed in this article uses as a measure of independence the mutual information between $Y$ and $Z$. This is not the only possible choice: other quantifiers of independence were briefly described in section \ref{sec:Formulation}, and still others are under development. The article used the numerical procedure developed to illustrate the broad applicability of the hierarchical barycenter concept through both real and synthetic examples.

\begin{appendix}
\section{Gradient of the objective function}
This section describes an alternative argument to the one developed in section \ref{sec:minimization}, for why one needs only consider the derivatives with respect to the first argument of the kernel functions. We focus again for brevity on the second term of (\ref{eq:R}) and consider its derivative with respect to $y_{l}$
\begin{equation}
\frac{\partial}{\partial y_{l}} \sum_{i\in N_{k}}\log\left(  \frac{1}{N_{k}} \sum_{j\in I_{k}} K^y(y_i,y_{j})\right)= \frac{\partial}{\partial y_{l}} \left (\log(\rhot(y_{l})) + \sum_{i\neq l} \log(\rhot(y_{i}))  \right)
\end{equation}
where for simplicity we wrote 
$$
\rhot(y_{l})= \frac{1}{N_{k}} \sum_{j\in I_{k}} K^y(y_l,y_{j}).
$$
Then
\begin{multline}
\frac{\partial}{\partial y_{l}} \left (\sum_{i\neq l} \log(\rhot(y_{i}))  \right)=\sum_{i\neq l}
\frac{1}{\rhot(y_i)}\frac{1}{N_{k}} \frac{\partial}{\partial y_{l}} \sum_{j\in I_{k}} K^y(y_i,y_{j})\approx \\
\approx \frac{\partial}{\partial y_{l}} \int \frac{1}{\rhot(y)} K^{y}(y,y_{l})\rhot(y)dy=0
\end{multline}
where we used the fact that the sum over $i$ approximates the expected value over $y$ with density $\rhot$ and where the last  equality follows from the fact that the kernel integrates to 1 for every $y_{l}$.

\end{appendix}

\bibliographystyle{plain}      %
\bibliography{Hierarchical}

\end{document}